\documentclass[12pt]{article}
\pdfoutput=1

\usepackage{ulem}

\usepackage[utf8]{inputenc}
\usepackage[left=2.55cm, right=2.55cm, top=2.55cm, bottom=2.55cm]{geometry}
\usepackage{amsmath,amssymb,amsbsy}
\usepackage{slashed}
\usepackage{xcolor}
\usepackage{graphicx}
\usepackage{url}
\usepackage{cancel}
\usepackage{cite}
\usepackage[colorlinks=true,allcolors=darkpurple,pdfborder={0 0 0},linktocpage=false]{hyperref}
\usepackage{tabularx,booktabs}
\usepackage{multicol}
\usepackage{units}
\usepackage{xspace}
\usepackage[labelfont=bf]{caption}
\usepackage[section]{placeins}
\usepackage{subcaption}
\usepackage{soul} 
\usepackage{changepage}

\definecolor{darkred}{rgb}{0.6,0,0}
\definecolor{darkpurple}{rgb}{0.5,0,0.5}

\def\hc{\text{h.c.}}

\def\z2{$\mathbb{Z}_2$}
\def\321{$\mathrm{SU(3)_c} \times \mathrm{SU(2)_L} \times \mathrm{U(1)_Y}$}
\def\one{\ensuremath{\mathbf{1}}}
\def\two{\ensuremath{\mathbf{2}}}
\def\three{\ensuremath{\mathbf{3}}}

\definecolor{avblue}{rgb}{0.0, 0.0, 0.8}

\newcommand{\AddrRic}{%
  Institut f\"ur Theoretische Physik und Astrophysik, Universit\"{a}t W\"{u}rzburg, 97074 Würzburg, Germany}

\newcommand{\AddrIFIC}{%
  Instituto de F\'{i}sica Corpuscular, CSIC-Universitat de Val\`{e}ncia, 46980 Paterna, Spain}

\newcommand{\AddrFISTEO}{%
  Departament de F\'{\i}sica Te\`{o}rica, Universitat de Val\`{e}ncia, 46100 Burjassot, Spain}

\begin{document}

\vspace*{-2cm}
\begin{flushright}
IFIC/22-23 \\
\vspace*{2mm}
\end{flushright}

\begin{center}
\vspace*{15mm}

\vspace{1cm}
{\Large \bf 
Neutrino masses, flavor anomalies and muon $\boldsymbol{g-2}$ from dark loops
} \\
\vspace{1cm}

{\bf Ricardo Cepedello$^{\text{a}}$, Pablo Escribano$^{\text{b,c}}$, Avelino Vicente$^{\text{b,c}}$}

\vspace*{.5cm}
 $^{(\text{a})}$ \AddrRic \\\vspace*{.2cm} 
 $^{(\text{b})}$ \AddrIFIC \\\vspace*{.2cm} 
 $^{(\text{c})}$ \AddrFISTEO

 \vspace*{.3cm} 
\href{mailto:ricardo.cepedello@physik.uni-wuerzburg.de}{ricardo.cepedello@physik.uni-wuerzburg.de}, \href{mailto:pablo.escribano@ific.uv.es}{pablo.escribano@ific.uv.es}, \href{mailto:avelino.vicente@ific.uv.es}{avelino.vicente@ific.uv.es}
\end{center}

\vspace*{10mm}
\begin{abstract}\noindent\normalsize
The lepton sector of the Standard Model is at present haunted by several intriguing anomalies, including an emerging pattern of deviations in $b \to s \ell \ell$ processes, with hints of lepton flavor universality violation, and a discrepancy in the muon anomalous magnetic moment. More importantly, it cannot explain neutrino oscillation data, which necessarily imply the existence of non-zero neutrino masses and lepton mixings. We propose a model that accommodates all the aforecited anomalies, induces neutrino masses and provides a testable dark matter candidate. This is achieved by introducing a dark sector contributing to the observables of interest at the 1-loop level. Our setup provides a very economical explanation to all these open questions in particle physics and is compatible with the current experimental constraints.
\end{abstract}

\section{Introduction}
\label{sec:intro}

Several anomalies are currently hinting at the presence of new physics
effects in the lepton sector of the Standard Model (SM). First of all,
and perhaps most importantly, neutrino oscillation experiments have
clearly established that neutrinos are massive. This is arguably the
most robust evidence of the existence of New Physics (NP) beyond the
SM. It definitely calls for an extension of the SM lepton sector with
new degrees of freedom which, in most scenarios, lead to deviations in
other observables, directly associated to leptons or
not. Interestingly, other anomalies have recently showed up, mainly
involving the muon:
\begin{itemize}
\item The hints observed in $b \to s$ transitions, which already point
  towards an emerging
  pattern~\cite{Altmannshofer:2021qrr,Hurth:2021nsi,Alguero:2021anc}. This
  includes a deviation in the branching ratio Br$(B_s \to
  \mu\bar\mu)$~\cite{LHCb:2015wdu,LHCb:2017rmj,ATLAS:2018cur} and,
  especially, the possible violation of lepton flavor universality in
  B-meson decays~\cite{LHCb:2017avl, BELLE:2019xld, Belle:2019oag,
    LHCb:2021trn} encoded in the $R_{ K^{(*)} }$ ratios, defined as
\begin{equation} \label{eq:RK_def}
    R_{ K^{(*)} } = \frac{ \text{Br}(B \to K^{(*)} \mu \bar\mu) }{ \text{Br}(B \to K^{(*)} e \bar e) } \, .
\end{equation}
For these ratios, the theoretical uncertainties of the SM predictions
are at the percent level~\cite{Bordone:2016gaq}, which strengthens the
relevance of the anomalies.
\item The muon anomalous magnetic moment $a_\mu = (g-2)_\mu /2$ has been recently measured with unprecedented accuracy~\cite{2104.03281} and in agreement with previous measurements from the E821 experiment at Brookhaven~\cite{hep-ex/0602035}. The combination of both observed values yields a deviation of $4.2\sigma$ from the SM predictions~\cite{2006.04822},\footnote{One should note, however, that the SM prediction is currently under debate due to some recent lattice results that weaken the anomaly~\cite{Borsanyi:2020mff,Ce:2022kxy,Alexandrou:2022amy}.}
  \begin{equation} 
      \Delta a_\mu = a_\mu^{\text{exp}} - a_\mu^{\text{SM}} = (2.51 \pm 0.59) \times 10^{-9} \, .
  \end{equation}
\end{itemize}
Finally, the nature of the dark matter (DM) that constitutes $\sim
25\%$ of the energy content of the Universe is still a mystery. Many
new physics models include DM candidates, sometimes relating them to
other open questions in particle physics or even being instrumental in
their resolution. While these NP indications might have different
origins, and some of them are still hints to be confirmed with further
experimental data and improved theoretical calculations, it is
tempting to consider a common explanation.

In this letter we introduce an economical, yet powerful model that
provides an explanation to all these new physics indications. This is achieved
thanks to the addition of a dark sector composed by a fermion singlet
$N$, two generations of inert doublets $\eta$, a doublet leptoquark
$S$ and a singlet scalar $\phi$, with the quantum numbers under
$\left( \rm SU(3)_c , \rm SU(2)_L \right)_{\rm U(1)_Y}$
\begin{equation}
  N \sim (\one,\one)_0 \, , \quad \eta \sim (\one,\two)_{\frac{1}{2}} \, , \quad S \sim (\three,\two)_{\frac{1}{6}} \, , \quad \phi \sim (\one,\one)_{-1} \, .
\end{equation}
The model also includes a dark \z2 parity, under which all the new
fields are assumed to be odd, while the SM fields are even. This
characterizes the dark sector of the model. As will be shown below,
these ingredients are enough to induce neutrino masses, accommodate
the $b \to s \ell \ell$ and muon $g-2$ anomalies, and provide a viable
DM candidate, while being compatible with all the relevant
experimental constraints. Therefore, our economical scenario takes
into account all the unresolved issues in the lepton sector
\textit{simultaneously} and, as a by-product, also addresses the
long-standing DM problem. In our model, all NP contributions to the
observables of interest are induced at the 1-loop level, with
\z2-odd particles running in the loop. These \textit{dark loops}
characterize our setup.

The connection between neutrino masses and the anomalies in $b \to s$
transitions has been explored in several works. In most cases,
neutrino masses are generated radiatively with one or several
leptoquarks participating in the loop. These leptoquarks are then
responsible for explaining at tree-level the flavor
anomalies~\cite{Pas:2015hca, Saad:2020ucl, Li:2022chc}, as well as the
muon $g-2$~\cite{Popov:2016fzr, Bigaran:2019bqv, Saad:2020ihm,
  Chang:2021axw, Nomura:2021oeu, Julio:2022bue, Chowdhury:2022dps,
  Chen:2022hle}. Ref.~\cite{Freitas:2022gqs} proposes an explanation
for the $b \to s \ell \ell$ anomalies via loops, also linked to the
generation of neutrino masses, while Ref.~\cite{Ashry:2022maw} takes a
completely different approach explaining the flavor anomalies in a
left-right model with neutrino masses generated through an inverse
seesaw. Finally, the $b \to s$ anomalies have also been discussed in
connection to the dark matter problem. We would like to highlight
Refs.~\cite{Huang:2020ris, Arcadi:2021cwg, Becker:2021sfd,
  Capucha:2022kwo}, which also address the $b \to s \ell \ell$ and
muon $g-2$ anomalies via loops involving the dark matter particle, and
refer to the review~\cite{Vicente:2018xbv} for other works in this
direction.

\section{The model}
\label{sec:model}

The new states in our model allow us to write the additional
Lagrangian terms:
\begin{align} 
  -\mathcal{L}_{\rm NP} &=
  Y_N \, \overline{N} \, \ell_L \, \eta
  + Y_S \, \overline{q}_L \, S \, N
  + \kappa \,  \overline{N^c} \, e_R \, \phi^\dagger
  + \frac{1}{2} \, M_N \, \overline{N^c} \, N
  + \hc \, . \label{eq:LYNP}
\end{align}
Here $Y_N$ is a $3\times 2$ matrix,  $Y_S$ and $\kappa$ are both $3$-components Yukawa vectors, while $M_N$ is a parameter with
dimensions of mass. Additional Yukawa couplings not written here are
forbidden by the dark \z2 parity. For instance, this is the case of the
$\overline{N} \, \ell_L \, H$ or $\overline{d}_R \, \ell_L \, S$
terms. The scalar potential of the model also contains many new terms,
including two that will be relevant for the discussion:
\begin{equation} \label{eq:VNP}
  \mathcal V_{\rm NP} \supset \frac{\lambda_5}{2} \, \left(H^\dagger \, \eta\right)^2 + \mu \, H \, \eta \, \phi + \hc \, .
\end{equation}
We remind the reader that two $\eta$ doublets are added to the field
inventory of the model. Therefore, $\mu$ is a $2$-component vector,
while $\lambda_5$ is a $2 \times 2$ symmetric matrix. In the
following, only the SM scalar doublet $H$ will be assumed to acquire a
non-zero vacuum expectation value (VEV), $H^0 = v/\sqrt{2}$, where $v
\simeq 246$ GeV is the usual electroweak VEV. This preserves the \z2
dark parity.

\paragraph{Neutrino masses}

\begin{figure}[t]
  \centering
  \includegraphics[width=0.4\linewidth]{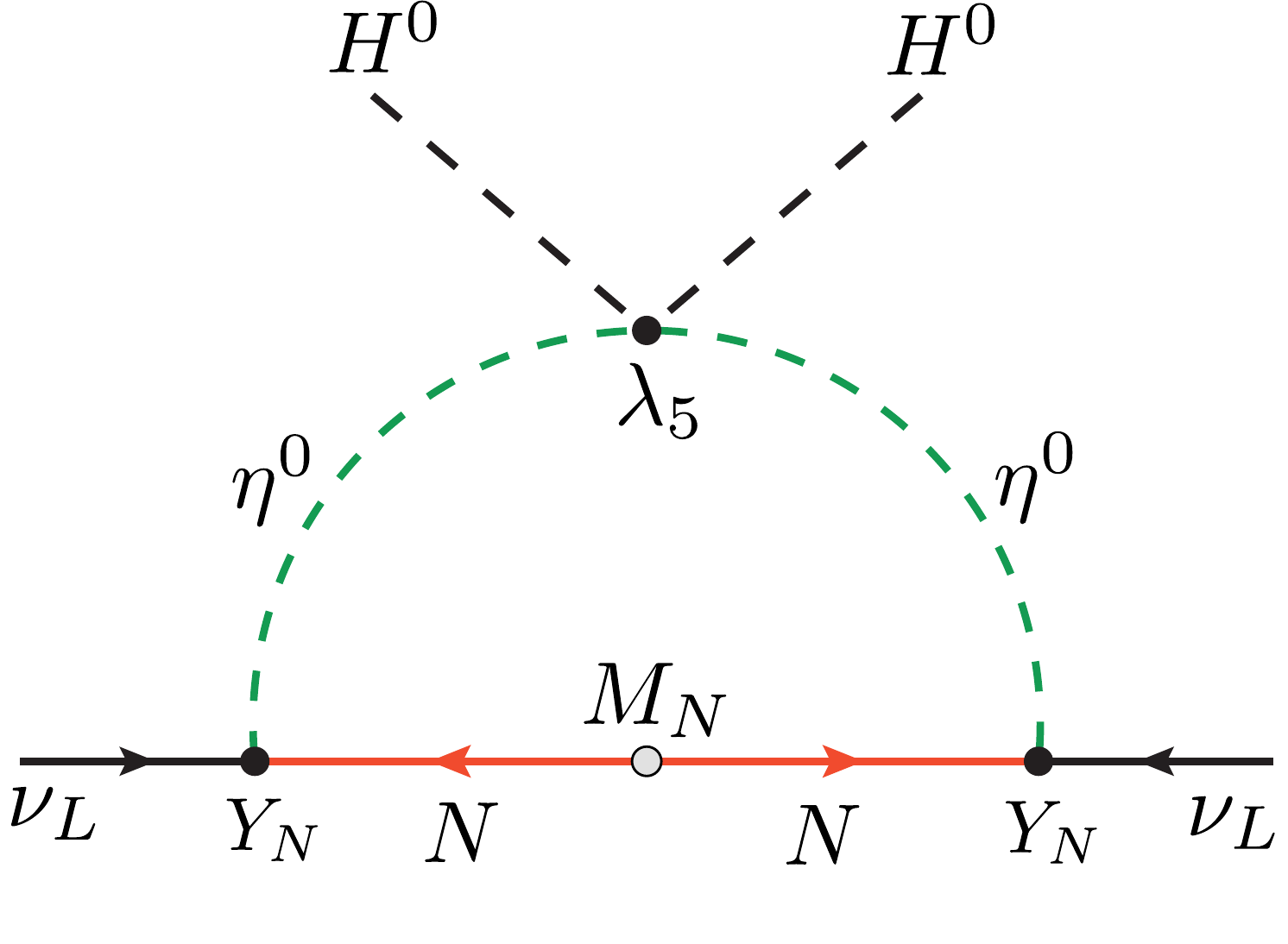}
  \caption{Generation of neutrino masses at the 1-loop level. In this
    diagram, $\eta^0$ represents the real and imaginary components of
    all generations of $\eta^0$.
  \label{fig:numass}
  }
\end{figure}

The conservation of \z2 prevents the generation of neutrino masses at
tree-level. However, the simultaneous presence of the $Y_N$, $M_N$ and
$\lambda_5$ terms in Eqs.~\eqref{eq:LYNP} and \eqref{eq:VNP} implies
the explicit breaking of lepton number in two units. Majorana neutrino
masses are induced at the 1-loop level \textit{a la
  Scotogenic}~\cite{Ma:2006km}, as shown in Fig.~\ref{fig:numass}. The
states running in the loop belong to the dark sector, a feature that
is enforced by the \z2 symmetry and will be common to all NP
contributions discussed below. The resulting neutrino masses in the
limit of small $\lambda_5$ is given by~\cite{Escribano:2020iqq}
\begin{equation} \label{eq:Mnu}
    (m_\nu)_{\alpha\beta} \approx \frac{1}{32 \pi^2} v^2 \, \sum_{a,b} (Y_N)_{\alpha a} (Y_N)_{\beta b} \, \lambda_5^{ab} \, \frac{ M_N }{m_b^2-M_N^2} \left[ \frac{m_b^2}{m_a^2-m_b^2} \log \frac{m_a^2}{m_b^2} - \frac{M_N^2}{m_a^2-M_N^2} \log \frac{m_a^2}{M_N^2} \right],
\end{equation}
%
%
%
were $m_{a,b}$ are the masses of the two $\eta$ doublets. This can be roughly estimated to be $m_\nu \sim \frac{\lambda_5 \, Y_N^2 \,
  v^2}{16 \pi^2 \, M_N}$. Therefore, $m_\nu \sim 0.1$ eV can be
obtained for $M_N = 1$ TeV and $\lambda_5^{ab} \sim
10^{-10}$, if the entries of the $Y_N$ matrix are of order $1$. The
smallness of the $\lambda_5$ elements is technically natural and
protected against radiative corrections~\cite{tHooft:1979rat}, since
in the limit $\lambda_5^{ab} \to 0$ lepton number is
restored. 

\section{Observables}
\label{sec:obs}

We now proceed to discuss the NP contributions induced by the new
states in our model to the observables of interest.

\begin{figure}[t]
  \centering
  \includegraphics[width=0.3\linewidth]{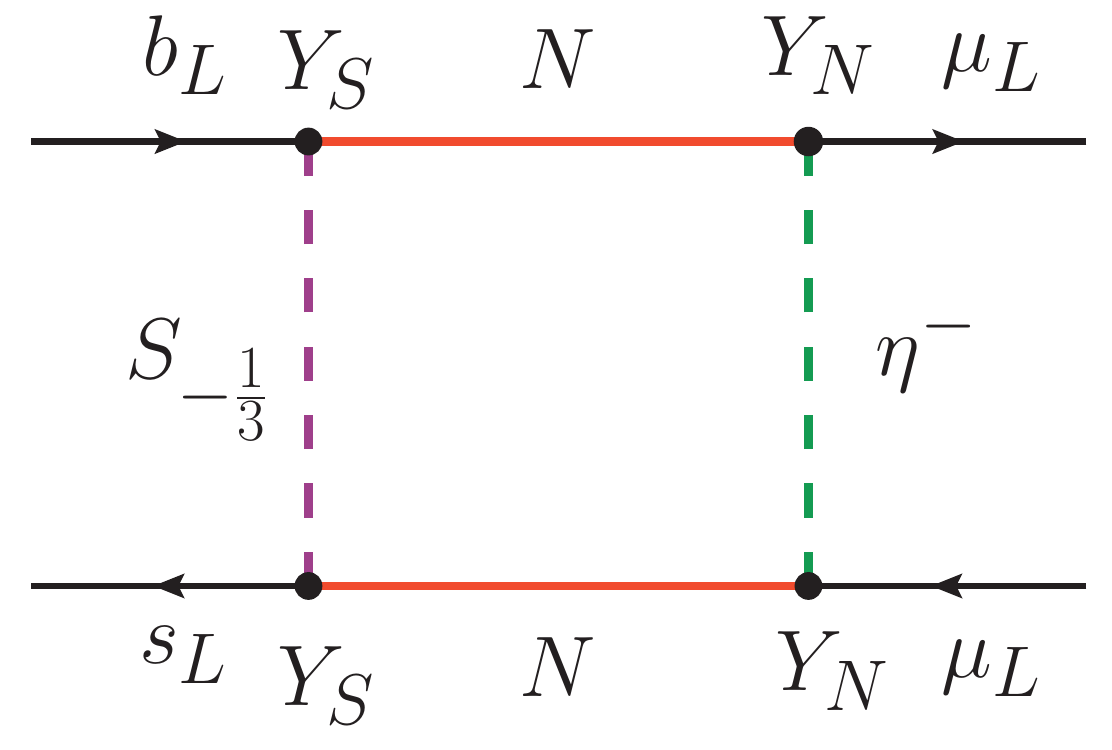}
  \includegraphics[width=0.3\linewidth]{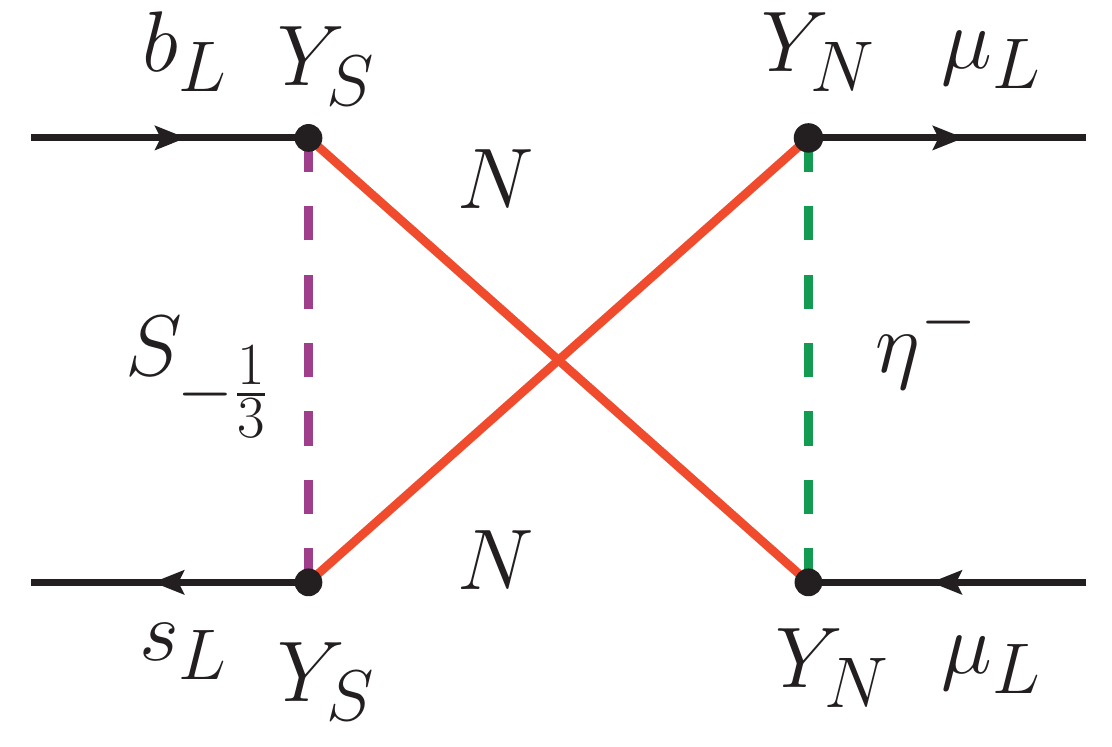}
  \includegraphics[width=0.3\linewidth]{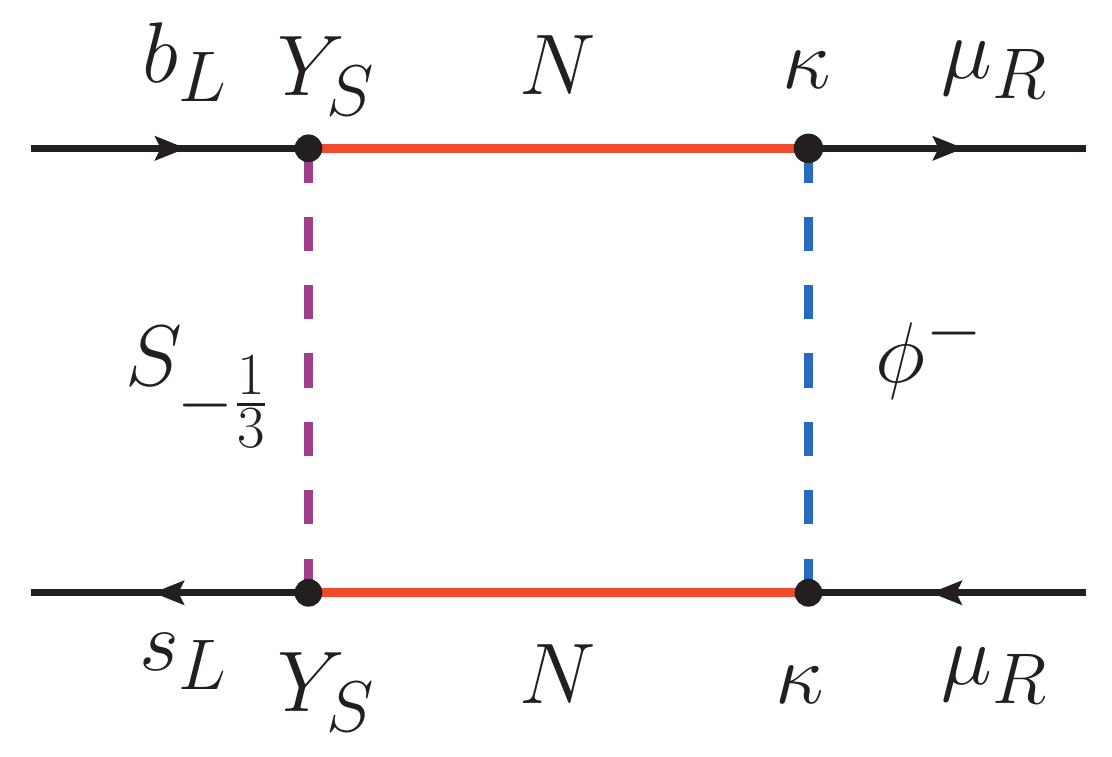} \hfill \\
  \includegraphics[width=0.35\linewidth]{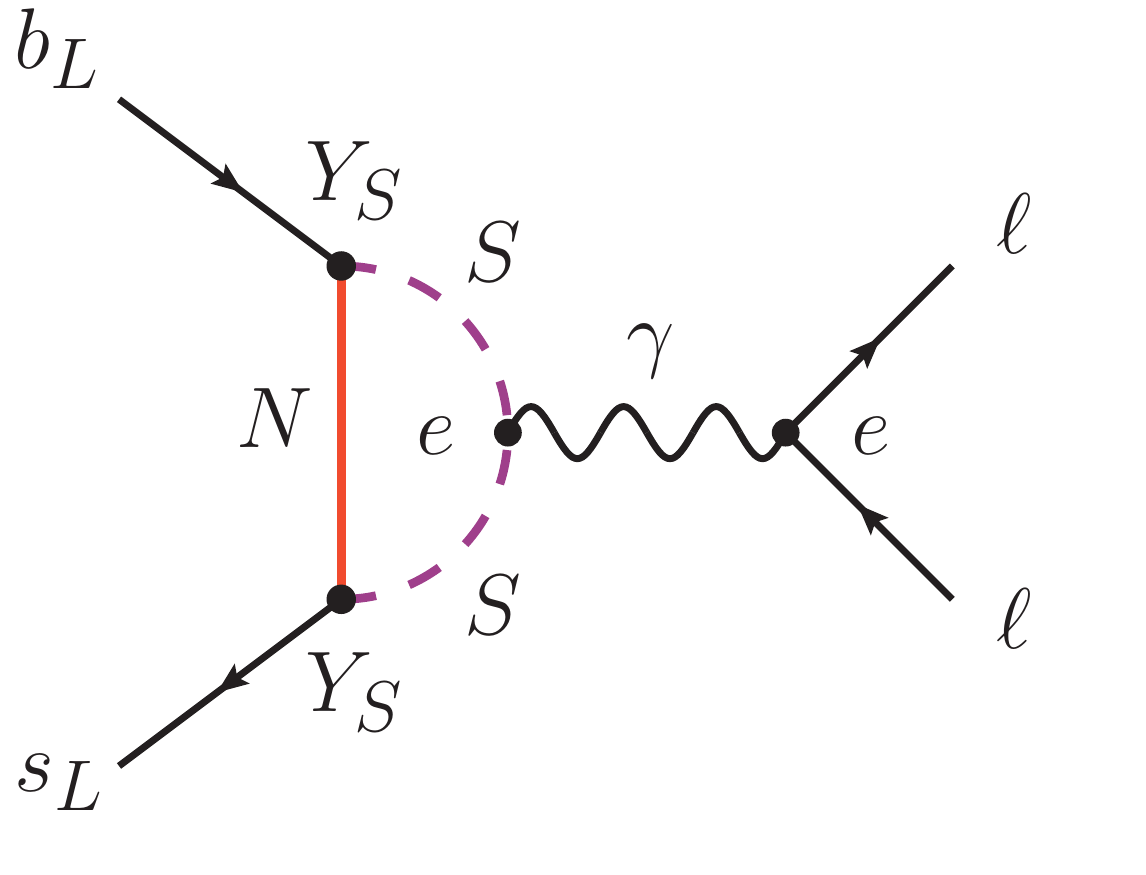}
  \hspace*{1cm}
  \includegraphics[width=0.3\linewidth]{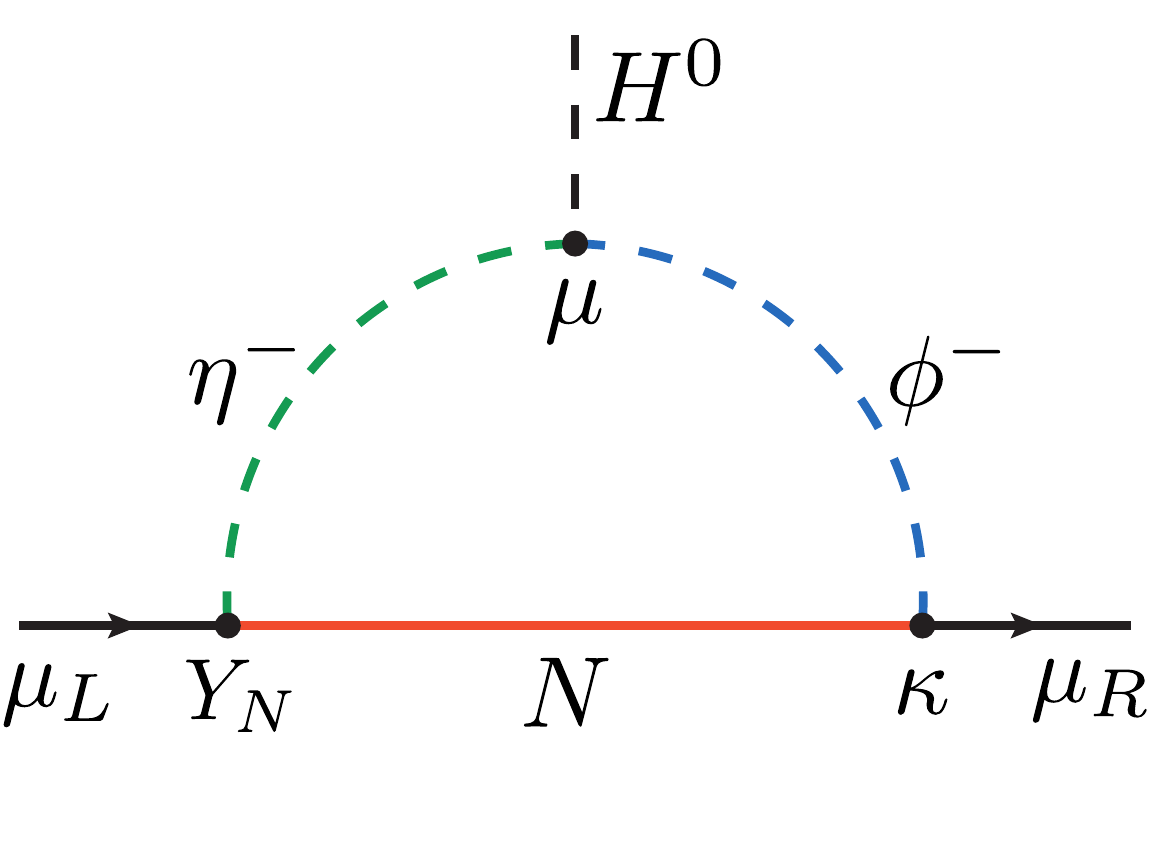}
  \caption{Some NP contributions to the observables of
    interest. Above: Box diagrams contributing to $b \to s \mu \mu$
    observables. Below left: Flavor-universal penguin contribution to
    the $b \to s \ell \ell$ anomalies. Below right: 1-loop
    contribution to the anomalous magnetic moment of the muon (the
    photon line should be attached to the charged scalars in the
    loop).
  \label{fig:NPcont}
  }
\end{figure}

\paragraph{$\boldsymbol{b \to s \ell \ell}$ anomalies}

The model induces many 1-loop contributions to $b \to s \ell \ell$
observables. Some examples of them are shown in
Fig.~\ref{fig:NPcont}. We note that the charged $\eta^-$ scalars mix
with the charged singlet $\phi^-$, and thus the states propagating in
the loops are the mass eigenstates resulting from this mixing. We
nevertheless show gauge eigenstates in Fig.~\ref{fig:NPcont} to better
illustrate the most relevant contributions. Box diagrams are
responsible for flavor universality violating contributions, central
to explain observables such as the $R_K$ ratio. In addition, one
should also take into account flavor universal contributions from
penguin diagrams, as shown in the second row of
Fig.~\ref{fig:NPcont}. Our model realizes scenario b)
of~\cite{Arnan:2016cpy}. We highlight the presence of the crossed
diagram in Fig.~\ref{fig:NPcont}, possible due to the Majorana nature
of the $N$ singlets. We do not show an analogous crossed diagram with
$\phi^-$ in the loop. These diagrams play a crucial role in canceling
unwanted contributions to $B_s - \overline{B}_s$
mixing~\cite{Arnan:2019uhr}, as discussed below.

\paragraph{Anomalous magnetic moment of the muon}

The model also has new contributions to the anomalous magnetic moment
of the muon, as shown in the lower row of Fig.~\ref{fig:NPcont}. One
should note that two different Yukawa couplings enter this
diagram. While $Y_N$ plays a role in the generation of neutrino
masses, the $\kappa$ Yukawa couplings do not. We also highlight the
presence of the $\mu$ trilinear couplings that induce mixing between
the charged component of $\eta$ and $\phi$ and can be used to
  chirally enhance the associated contribution.

\paragraph{Dark matter}

Last but not least, the model also provides a solution to the DM
problem. The lightest \z2-odd state is stable and, if electrically
neutral, it is a potentially valid DM candidate. Two possibilities
arise: the lightest $N$ state and one of the components (CP-even or
CP-odd) of the neutral $\eta^0$ scalars. Both scenarios have been
widely studied in the literature for the pure Scotogenic model~\cite{Ma:2006km} and
both have been shown to be compatible with the observed DM relic
density. However, we note that the scalar candidate can achieve this
more easily~\cite{Deshpande:1977rw,Barbieri:2006dq,LopezHonorez:2006gr,Honorez:2010re,Diaz:2015pyv}, since the fermionic candidate requires large $Y_N$ Yukawa
couplings and then leads to some tension with existing bounds from
lepton flavor violating observables~\cite{Vicente:2014wga}.

\section{Numerical results}
\label{sec:results}

Our model faces several experimental constraints. First of all, we
must make sure that neutrino oscillation data are correctly
reproduced. We use the results of the global
fit~\cite{deSalas:2020pgw} and implement them by means of a
Casas-Ibarra parametrization~\cite{Casas:2001sr}, properly adapted to
the Scotogenic
scenario~\cite{Toma:2013zsa,Cordero-Carrion:2018xre,Cordero-Carrion:2019qtu}. This
allows us to write the $Y_N$ Yukawa matrix as
\begin{equation} \label{eq:CI}
Y_N^T = V \, D_{\sqrt{\Sigma}} \, R \, D_{\sqrt{m_\nu}} \, U^\dagger_{\text{PMNS}} \, ,
\end{equation}
where $R$ is a general $2 \times 3$ orthogonal
matrix defined as
\begin{equation}
	R = \left( \begin{matrix} 0 & \cos \theta & -\sin \theta \\
                                  0 & \sin \theta & \cos \theta \end{matrix} \right) \, .
\end{equation} 
$D_X$ represents the diagonal form of the matrix $X$, while $\Sigma$
is defined from Eq.~\eqref{eq:Mnu} as $m_{\nu} = Y_N \cdot \Sigma
\cdot Y_N$, with $V$ its diagonalization matrix. Finally,
$U_{\text{PMNS}}$ is the usual unitary matrix that relates neutrino
flavor to neutrino mass eigenstates. Eq.~\eqref{eq:CI} illustrates an
important connection in our model: neutrino masses strongly restrict
the elements of the $Y_N$ Yukawa matrix, which play a crucial role in
the resolution of the $b \to s \ell \ell$ and $(g-2)_\mu$ anomalies
(see Fig.~\ref{fig:NPcont}). Furthermore, as in most neutrino mass
models, lepton flavor violating processes, such as $\mu \to e \gamma$,
are potentially dangerous. Searches by the MEG collaboration have
shown that the branching ratio for this radiative decay cannot exceed
$4.2 \times 10^{-13}$~\cite{MEG:2016leq}.

Regarding processes with mesons, the main constraints come from $b
  \to s \gamma$, $B \to K^{(*)} \nu \bar \nu$ and $B_s -
\overline{B}_s$ mixing. The $b \to s \gamma$ decays yield strong
  constraints on the coefficients of dipole
  operators~\cite{Misiak:2020vlo}. These are induced at the 1-loop
  level by diagrams like the one shown in the lower left corner of
  Fig.~\ref{fig:NPcont}, with photons or gluons and without the
  charged leptons. The inclusive $b \to s \gamma$ branching ratio is
  experimentally determined to be BR$(b \to s \gamma) = (3.49 \pm
  0.19) \times 10^{-4}$~\cite{HFLAV:2022pwe}. As we will show below,
  this constrains the $\left(Y_S\right)_{2} \times
  \left(Y_S\right)_{3}$ product. About $B \to K^{(*)} \nu \bar \nu$, note that if a contribution to $B
\to K^{(*)} \ell^+ \ell^-$ exists, the corresponding process with neutrinos is
unavoidable due to $\rm SU(2)_L$ invariance. Current experimental
results set limits to the branching ratios of $B \to K^{(*)} \nu \bar
\nu$ which, normalized to their SM predictions, are restricted to
$R_{K}^{\nu\bar\nu} < 3.9$ and $R_{K^*}^{\nu\bar\nu} <
2.7$~\cite{Belle:2017oht}. On top of this, $B_s - \overline{B}_s$
mixing~\cite{ParticleDataGroup:2018ovx} is again inevitable and
typically very constraining in scenarios aiming at an explanation of
the $b \to s \ell \ell$ anomalies at the 1-loop level. Any model that
generates a box diagram with $b$ and $s$ quarks and two leptons, like
the ones in Fig.~\ref{fig:NPcont}, will automatically produce box
contributions to the four quarks operators responsible for $B_s -
\overline{B}_s$ mixing. In fact, this specific constraint precludes
most radiative models for $b \to s$ transitions. In our scenario,
however, the Majorana nature of the $N$ singlets can be used to
suppress $B_s - \overline{B}_s$ mixing in the limit of (nearly)
degenerate NP masses participating in the box, i.e. $S_{-1/3}$ and
$N$, as pointed out in~\cite{Arnan:2019uhr}.

We present now our results. Our goal is to prove that our model can
accommodate all the anomalies while being consistent with neutrino
oscillation data and all the experimental constraints. In what
concerns the $b \to s \ell \ell$ anomalies, a reasonable goal is to
accommodate Scenario 5 of the global fit~\cite{Alguero:2021anc},
characterized by
\begin{equation} \label{eq:bfp}
  \begin{gathered}
    \mathcal{C}_{9\mu}^V = -0.55^{+0.44}_{-0.47} \, , \\
    \mathcal{C}_{10\mu}^V = 0.49^{+0.35}_{-0.41} \, , \\
    \mathcal{C}_9^U = \mathcal{C}_{10}^U = -0.35^{+0.42}_{-0.38} \, ,
  \end{gathered}
\end{equation}
where $\mathcal{C}_9$ and $\mathcal{C}_{10}$ are the Wilson
coefficients of the $\mathcal{O}_9 = (\bar s \gamma_\mu P_L b)(\bar
\ell \gamma^\mu \ell)$ and $\mathcal{O}_{10} =(\bar s \gamma_\mu P_L
b)(\bar \ell \gamma^\mu \gamma_5 \ell)$ effective operators,
respectively. The superindices $V$ and $U$ denote flavor universality
violating and conserving contributions, respectively, and the flavor
universality violating ones are specific to the muon flavor. This
  scenario provides a clear improvement with respect to the SM in what
  concerns the description of $b \to s \ell \ell$
  data~\cite{Alguero:2021anc}.
We constructed a $\chi^2$ function in the usual way, with these four
Wilson coefficients and the anomalous magnetic moment of the
muon. Rather than finding the global minimum of the resulting $\chi^2$
function, which depends non-trivially on many model parameters, our
goal is to prove that our model can provide a good explanation to all
the anomalies. Therefore, in order to simplify the analysis, we fixed
several parameters. First of all, the masses of the NP states were
taken to be close to $1$ TeV, a typical reference NP scale. We have
explicitly checked that the qualitative results and the conclusions of
our analysis remain the same with other choices of NP scale. Note that
the $B_s - \overline{B}_s$ mixing suppression requires the masses of
$S$ and $N$ to be degenerate or nearly
degenerate~\cite{Arnan:2019uhr}. The mass of $\eta$ is taken to be
lower, around $550$ GeV. With this hierarchy, $\eta$ would be the
lightest stable particle with a mass compatible with the observed DM
relic density and direct detection cross-section
bounds~\cite{Honorez:2010re,Diaz:2015pyv}. We assumed that the
$2\times 2$ matrix $\lambda_5$ is proportional to the identity,
i.e. $\lambda_5 = \lambda_5^0 \, \mathbb{I}_2$, and fix the following
values:
\begin{align}
    \mu_{1} = -\mu_{2} &= - 1.0 \, \text{TeV} \, ,& \kappa_{1} &= 0 \, ,
    \\
    \lambda_5^0 &= 2 \times 10^{-10} \, ,&\kappa_{2} &= 0.04 \, .
\end{align}
We noticed that both elements of the coupling vector $\kappa =
(\kappa_1 \; \kappa_2)$ need to be small in order to suppress the
branching ratio of $\mu \to e \gamma$ below the experimental
bound. Indeed, we chose to set $\kappa_{1}$ to zero. The rest of the
parameters of the model are not relevant for our discussion here,
given the generation structure of the diagrams depicted in
Fig.~\ref{fig:NPcont}. Note also that due to the external quark
structure, $Y_S$ always enters in the combination
$\left(Y_S\right)_{2} \times \left(Y_S\right)_{3}$. For the $\chi^2$
minimization, we are then left with $\left(Y_S\right)_{2} \times
\left(Y_S\right)_{3}$ and $\sin \theta$. We found that the values of
the parameters for which $\chi^2$ was minimal, were
\begin{equation}
  \left(Y_S\right)_{2} \times \left(Y_S\right)_{3} = 0.6 \, , \quad \sin \theta = 0.25 \, ,
\end{equation}
giving $\chi^2_{\text{min}} = 1.52$ and $\Delta \chi^2 =
  \chi^2_{\text{SM}} - \chi^2_{\text{min}} = 21.23$. This not only
  shows a remarkable improvement with respect to the SM, but the low
  $\chi^2_{\text{min}}$ value also guarantees that all the anomalous
observables can be properly accommodated in our model.
\begin{figure}[t]
  \centering
  \includegraphics[width=0.48\linewidth]{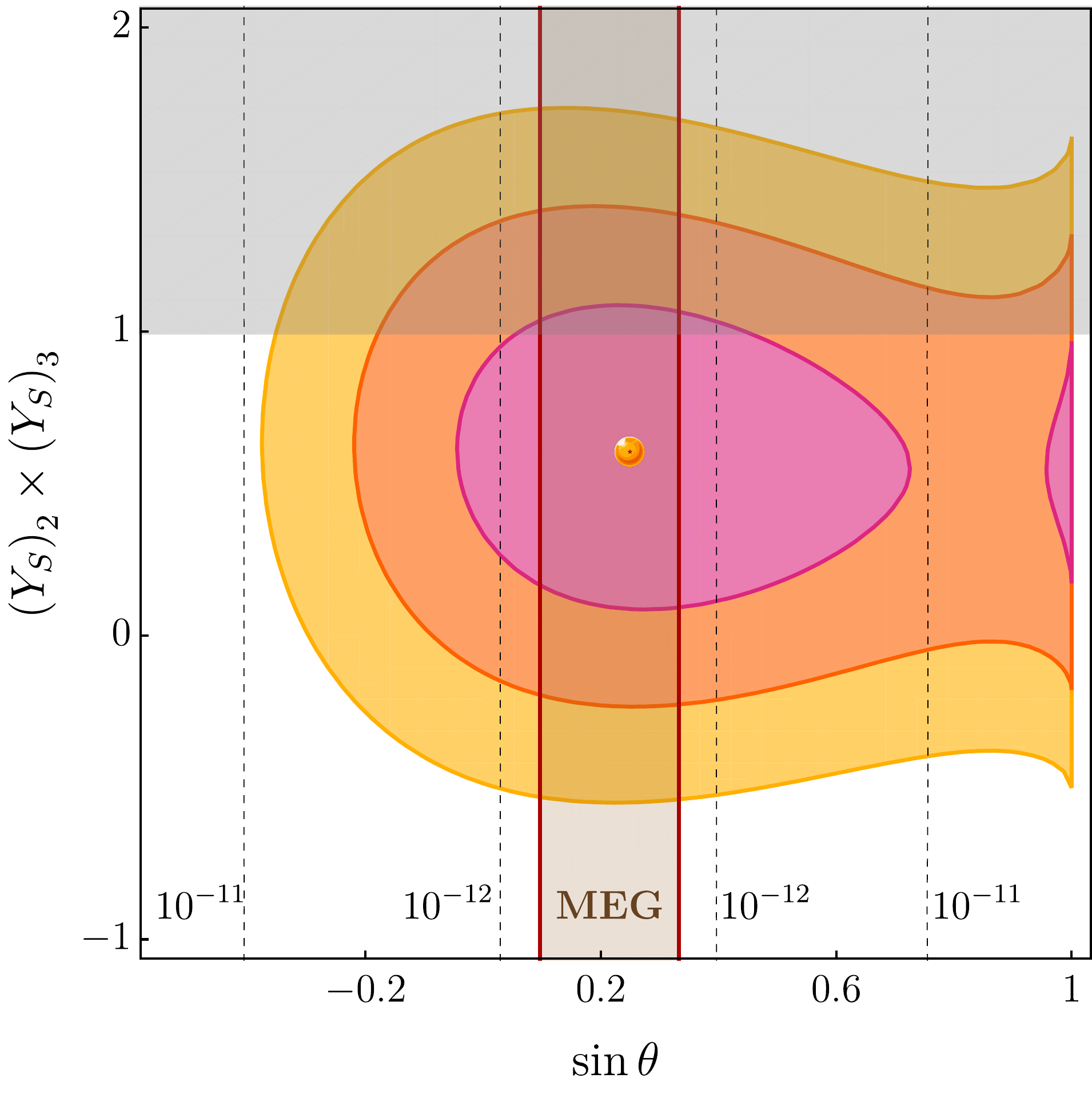}
  \hspace*{0.2cm}
  \includegraphics[width=0.48\linewidth]{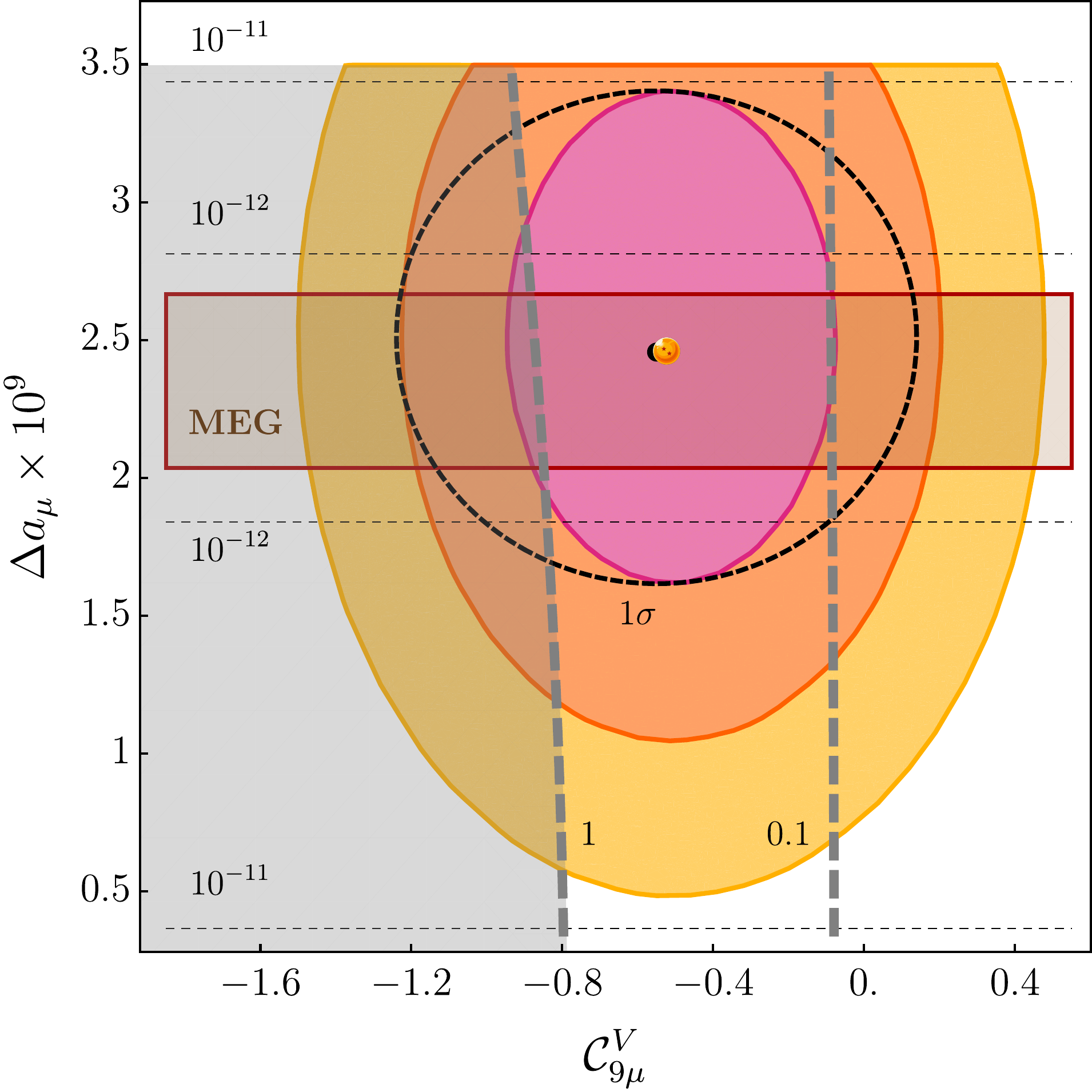}
  \caption{
Results of our $\chi^2$ fit of the model parameters. The colored
regions correspond to $1\,\sigma$ (pink), $2\,\sigma$ (orange) and
$3\,\sigma$ (yellow) regions, while the best-fit point is indicated
with an orange dot. The region allowed by the MEG experiment is shown
in dark red, while the dashed lines correspond to contours of Br($\mu
\to e \gamma$). The shaded region is excluded by the $b \to s
  \gamma$ constraint at $3 \, \sigma$. In the right panel, contours
of the $\left(Y_S\right)_{2} \times \left(Y_S\right)_{3}$ product are
shown with thick dashed gray lines, while the black dashed line and
dot are the experimentally determined $1\,\sigma$ region and central
value, respectively.
  \label{fig:results}
  }
\end{figure}
This is better illustrated on the left-hand side of
Fig.~\ref{fig:results}, which shows the results of our $\chi^2$ fit in
the $\sin \theta - \left(Y_S\right)_{2} \times \left(Y_S\right)_{3}$
plane. We find that both parameters can substantially deviate from
their best-fit values without affecting the $\chi^2$ function
notably. However, $\sin \theta$ is required to be in the $0.25$
ballpark in order to reduce the $\mu \to e \gamma$ branching ratio
below its current experimental bound. This turns out to be a strong
constraint in our model due to the connection to neutrino masses,
which generically require the $Y_N$ couplings involving the electron
to be non-zero. Similarly, the $b \to s \gamma$ constraint imposes
  an upper bound on the $\left(Y_S\right)_{2} \times
  \left(Y_S\right)_{3}$ product, which has to be below $\sim 1$.
The impact on $\mathcal{C}_{9\mu}^V$ and $\Delta a_\mu$ is shown on
the right-hand side of Fig.~\ref{fig:results}. Here we see that the
central value for both \textit{observables} (we treat the
$\mathcal{C}_{9\mu}^V$ coefficient as an observable here) can be
easily achieved in our model and is in fact very close to our global
best-fit point in Eq.~\eqref{eq:bfp}, which only deviates slightly due
to the influence of other Wilson coefficients. It is also remarkable
that our model does not require too large $Y_S$ Yukawa parameters to
accommodate the $b \to s \ell \ell$ anomalies. In fact,
$\mathcal{O}(1)$ $Y_S$ Yukawas are sufficient to reproduce all the
anomalies at the $1\,\sigma$ level. We emphasize once again that all
the parameter points considered in our analysis comply with the
constraints from neutrino oscillation data, $b \to s \gamma$ and $B_s
- \overline{B}_s$ mixing. The $B \to K^{(*)} \nu \bar \nu$ bounds are
also easily satisfied. Finally, the mass spectrum chosen in our
numerical fit also accommodates the observed DM relic density.

\section{Discussion}
\label{sec:conclusions}

Several anomalies have been found in recent years in observables
associated to leptons: a set of deviations in $b \to s \ell \ell$
observables and the muon $g-2$ discrepancy. In addition, there is the
long-standing problem of neutrino masses and mixings. This leaves us
with the well-motivated task of finding a simultaneous explanation to
all the NP indications in the lepton sector in a framework that is
also capable of generating naturally small neutrino masses.

We have proposed a novel model that accommodates all the aforementioned
anomalies, induces neutrino masses and provides a weakly-interacting
dark matter candidate, thanks to a dark sector including several
states contributing to the observables of interest at the 1-loop
level. We have shown that our simple and economical model can explain
all the anomalies via these dark loops. This is achieved with
renormalizable Yukawa couplings, while being compatible with neutrino
oscillation data and the existing experimental bounds. The flavor
violating muon decay $\mu \to e \gamma$ turns out to provide an
important constraint, but this can be easily satisfied in a wide
region of the parameter space of the model.

The scenario considered in our analysis requires the existence of
several states at the TeV scale. Since they are all odd under a new
dark parity, their production and decay channels are modified with
respect to more common scenarios. For instance, the $S$ leptoquark
must be produced in pairs at colliders and subsequently decay as $S
\to j \, N \to j \, \ell \, \slashed{E}_T$, where the jet can be given
by a 2nd or 3rd generation quark and the missing energy in the final
state is due to the production of the $\eta$ dark matter particle. The
approximate mass degeneracy between $S$ and $N$, introduced to
suppress $B_s - \overline{B}_s$ mixing, implies very soft jets,
undetectable at the LHC. Moreover, if both $\eta$ generations are
lighter than $N$, additional leptons can be produced in the cascade. A
more compressed spectrum would not affect our results in a
substantial way, but would make these leptons very soft too, leading to
a particularly challenging scenario at the LHC. In what concerns the
heavy neutral lepton $N$, the conservation of the dark parity forbids
its mixing with the standard neutrinos. We conclude that our scenario
contains several non-standard features and a dedicated study is thus
required to fully assess its observability. We leave it for future
work.

\section*{Acknowledgements}

Work supported by the Spanish grants PID2020-113775GB-I00
(AEI/10.13039/501100011033) and CIPROM/2021/054 (Generalitat
Valenciana). AV acknowledges financial support from MINECO through the
Ramón y Cajal contract RYC2018-025795-I. RC is supported by the
Alexander von Humboldt Foundation Fellowship. The work of PE is
supported by the FPI grant PRE2018-084599.



\bibliographystyle{utphys}
\bibliography{refs}

\end{document}